\documentclass{article}
\usepackage{graphicx}
\begin{document}

\title{An algorithm for quantum gravity phenomenology}
\author{Yuri Bonder\\
Instituto de Ciencias Nucleares\\
Universidad Nacional Aut\'onoma de M\'exico\\
Apartado Postal 70-543, Ciudad de M\'exico, 04510, M\'exico}

\date{bonder@nucleares.unam.mx}

\maketitle
\begin{abstract}
Quantum gravity phenomenology is the strategy towards quantum gravity where the priority is to make contact with experiments. Here I describe what I consider to be the best procedure to do quantum gravity phenomenology. The key step is to have a generic parametrization which allows one to perform self-consistency checks and to deal with many different experiments. As an example I describe the role that the Standard Model Extension has played when looking for Lorentz violation.
\end{abstract}

\section{Introduction}

Finding a fully consistent theory of quantum gravity is one of the biggest challenges in physics (for an introductory review see Ref.~\cite{QG}). One obstacle to build this theory is the lack of experimental data. The program generically known as quantum gravity phenomenology is comprised by several research strategies  sharing the willingness to make contact with experiments \cite{QGP}. Basically, the starting point for any of this strategies is to make an educated guess on how can a quantum gravity effect arise at testable scales. Then, some mathematical expressions are given containing free parameters that must be constrained by comparing with experiments and, for which, a nonzero value can be considered to be evidence of new physics.

\begin{figure}
\begin{center}
\includegraphics[width=\textwidth]{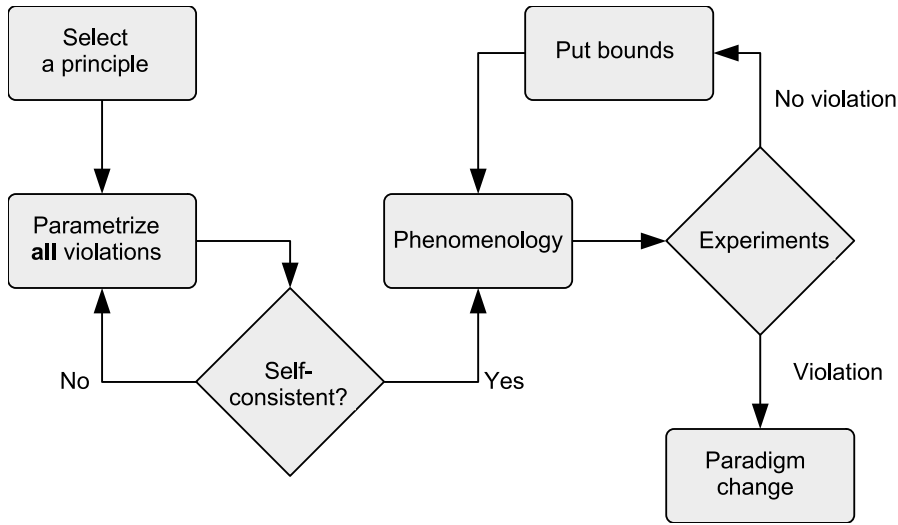}
\end{center}
\caption{\label{alg}The proposed algorithm for quantum gravity phenomenology.}
\end{figure}

The claim is that, by refining the methods that are usually utilized, we could pursue the goals of quantum gravity phenomenology more efficiently. I introduce the proposal as an algorithm whose flowchart is presented in figure \ref{alg}. I first describe this algorithm; then, as an example, I discuss how it has been implemented in the case of Lorentz violation.

\section{The algorithm}

The first step of the proposed algorithm is to narrow down the search for new physics. Of course, given that we know essentially nothing about the quantum gravity regime, there is no criteria that could help us make this decision. However, one can classify the possible research avenues. The proposal is to do so in terms of the principles of the current paradigm. That is, to select one of these principles and then look for deviations from conventional physics where this particular principle does not hold.

The use of principles as a starting point is a natural choice because, as happened in every scientific revolution, a dramatic change in this part of the theory must occur \cite{Kuhn}. Note, however, that this classification is not problem free. For example, there is no full consensus on what the principles underlying current physics are or how should they be enunciated. In addition, there could be some principles that are not independent. However, using the most basic structure in the theory minimizes the risks of propagating ambiguities into the following steps.

Once a principle is chosen it is extremely useful to parametrize all possible deviations from conventional physics where this principle is violated. Looking for \emph{all} modifications reflects the fact that we have no clues about the nature of quantum gravity. I would like to stress that the quantum gravity phenomenology literature is filled with examples where the opposite view point is taken.

Moreover, a generic parametrization can encode all relevant experimental results as conditions on its parameters. In this sense the parametrization provides us with a tool to compare different experiments. Even more, it can motivate new experiments since the mathematical expressions can guide our physical intuition, as often occurs in physics. Finally, the fact that the parametrization is done on top of conventional physics is useful to keep track of the physical meaning of the mathematical objects at hand.

It is important to stress that, in practice, it is hard to write this general parametrization. What typically happens is that one uses additional criteria to select a subspace of the parametrization and tries to be general under such a restriction. Something useful to define these subspaces is the fact that the structure of conventional physics permeates into the parametrization, providing a classification of all possible new effects in terms of `sectors.' For example, one can decide to focus only on an extension to Quantum Electrodynamics, or General Relativity, and so on. Moreover, the parametrization may have an intrinsic hierarchy based, essentially, on dimensional analysis, and which may be used to argue what modifications could dominate. Once the generic parametrization is built, it is possible to verify if it is internally consistent. This is a paramount step, which is often ignored, that can discard parts of the parametrization.

Finally, once the self-consistent parametrization is obtained, it is time to compare with experiments. For that purpose, and given a concrete experiment, it is useful to select the relevant sector and to perform the corresponding approximations. Clearly, the resulting expressions carry the parameters and, if no evidence of violations is shown, bounds on such parameters can be placed. In principle, one would then try to improve the experiment and continue testing, improving the limits on the parameters on each cycle. Of course, there is always the chance to detect a signal that leads to a paradigm change, which is the ultimate goal of the program.

It should be mentioned that, even if no signs of new physics are uncovered, the results of these analyses are extremely valuable. This is because it motivates new experiments, which can have unexpected outcomes, but, more importantly, since the bounds on the parameters can be used to rule out different theories and to have a quantitative measure of the validity of the principle that is tested.

\section{An example: Lorentz violation}

The community that best implements the method described above revolves around the so-called Standard Model Extension (SME), which is a parametrization for violations of local Lorentz invariance \cite{SME,Kostelecky2004}. Recall that local Lorentz invariance states that locally and in free fall there are no preferred directions associated with spacetime. Now, the SME is built in the effective field theory context. Thus, it has an action containing the conventional-physics action plus new terms that are not Lorentz invariant \cite{KosteleckyPotting}. These new terms have a parameter for Lorentz violation, called an SME coefficient, which is coupled to conventional fields. Also, the field content and the gauge symmetries of conventional physics are unchanged. Remarkably, the SME has triggered many theoretical and experimental works, some of which are described next.

\subsection{Self-consistency checks}

One of the earliest theoretical analysis done within the SME has to do with field redefinitions \cite{fieldredef}. It was noted that such redefinitions can be used to cancel some terms. These studies explained why some coefficients did not show up in the phenomenological expressions and are now a key part of any phenomenological analysis.

Another example is related to the fact that, when Lorentz violation is studied on a dynamical spacetime, the metric equation of motion contains the Einstein tensor, which is divergence free by virtue of the Bianchi identity. Thus, the divergence of this equation severely restricts the values of the SME coefficients. The most popular solution within the SME community is to consider that Lorentz violation arises spontaneously \cite{Kostelecky2004}. That is, that there are action terms for the coefficients with potentials that favor those configurations where preferred directions are selected.

There are methods to avoid using a particular action term for each SME coefficient \cite{Bailey}. However, these methods can only be implemented perturbatively and cannot be used to look for the effects of all the SME coefficients \cite{tpuzz}. Note in passing that gravitational effects, for tabletop experiments, can be introduced without having to use spontaneous Lorentz violation by using a uniformly accelerated frame \cite{Newtonian}.

The last example I want to mention has to do with the Cauchy problem. A desirable property of physical theories is that, given proper initial data, they make unique predictions. When the theories also satisfy particular causality and continuity relations between initial-data changes and solution changes, they are said to have a well-posed Cauchy problem \cite{Wald}. Now, the most popular models of spontaneous Lorentz violation, the so-called Bumblebee Models, consist of a vector field with a Maxwell-like kinetic term and potentials for the field square that generate spontaneous Lorentz violation \cite{Bumblebees}. It has been shown that these models, in flat spacetime and coupled to a scalar field that plays the role of matter, do have a Hamiltonian that respects the constraints \cite{BonderEscobar}. However, it was also shown that the initial data, as required by the Dirac analysis \cite{Dirac}, does not determine the evolution uniquely \cite{BonderEscobar}. Concretely, given a set of initial data, it is possible to find several physically-inequivalent evolutions that are all consistent with the same initial data. This is an example of a consistency analysis that can be used to discard some SME parts.

An easy exit to the objection described in the last paragraph is to simply change the kinetic term. The problem, however, is that most non-Maxwell kinetic terms depend on the spacetime connection. Therefore, by fixing the Cauchy problem for the vector field one may inadvertently damage that of the metric. This should be studied carefully but it is appealing as it can restrict further these models for spontaneous Lorentz violation.

\subsection{Experimental results}

The most exciting feature of the SME is the continuous flow of new experiments that look for Lorentz violation. These experiments are of various types; a partial list includes: 
\begin{itemize}
\item Accelerator and collider experiments.
\item Astrophysical observations.
\item Vacuum birefringence and dispersion of light.
\item Clock-comparison experiments.
\item Laboratory gravity tests.
\item Matter interferometry.
\item Neutrino oscillations.
\item Particle \textit{vs}.~antiparticle comparisons.
\item Resonant cavities and lasers.
\item Sidereal/annual variations of physical signals.
\end{itemize}

All the results from these experiments are collected in the so-called Data Tables for Lorentz Violation \cite{DataTables} that quote more that $150$ experimental results. Remarkably, several of the empirical limits that have been obtained are transplanckian. In fact, in some cases the limits approach two powers of the ratio of the electroweak and the Planck scales, which is the number where one can naively expect quantum gravity effects to show up.

\section{Conclusions}

There is an active community working on quantum gravity phenomenology for which the main goal is to search for experimental clues of quantum gravity. Here a procedure to work in the area is presented. The basic nonstandard steps are to have a generic parametrization of all possible violations of a principle and to do self-consistency checks. This strategy has been implemented mainly by the Standard Model Extension program that has motivated many interesting experimental and theoretical studies.

It should be said that, thus far, there is no compelling empirical evidence of quantum gravity. This only means that there is a lot of work to be done and, ideally, many studies will follow the steps outlined here. I am confident that, if we keep testing for the principles of current physics, we will eventually uncover new physics.

\section*{Acknowledgements}
I thank Alan Kosteleck\'y for many discussion on this topic and I acknowledge financial support from UNAM-DGAPA-PAPIIT Grant No.~IA101116 and Red Tem\'atica CONACyT ``F\'isica de Altas Energ\'ias.''

\end{document}